%% file: main.tex
\documentclass[sigconf,screen]{acmart}

\usepackage{amsmath,amssymb,amsfonts}
\usepackage{algorithmic}
\usepackage{graphicx}
\usepackage{textcomp}
\usepackage{xcolor}
\usepackage{balance} 
\usepackage{url} 
\def\BibTeX{{\rm B\kern-.05em{\sc i\kern-.025em b}\kern-.08em
    T\kern-.1667em\lower.7ex\hbox{E}\kern-.125emX}}

\usepackage{textcomp}
\usepackage{booktabs}
\usepackage{enumitem}
\usepackage{multirow}
\usepackage{subfig}
\usepackage{listings}
\usepackage{color}
\usepackage{threeparttable}
\usepackage{array}
\usepackage{diagbox}
\usepackage{graphicx}
\usepackage{verbatim}
\usepackage{url}
\usepackage{soul}
\soulregister\cite7
\soulregister\ref7
\usepackage[multiple]{footmisc}
\usepackage{graphicx}
\usepackage{multirow}
\usepackage{hhline}
\usepackage{comment}
\usepackage{float}
\usepackage{array,booktabs}
\usepackage{marvosym} %
\usepackage{pifont}
\usepackage{tcolorbox}
\usepackage{bm}
\usepackage[ruled,linesnumbered]{algorithm2e}

\newcommand{\tabincell}[2]{\begin{tabular}{@{}#1@{}}#2\end{tabular}}

\newcommand{\find}[1]{
\begin{tcolorbox}[leftrule=0.5mm,rightrule=0.5mm, toprule=0.5mm,bottomrule=0.5mm,left=2pt,right=2pt,top=2pt,bottom=2pt]%
\em #1
\end{tcolorbox}
}

\newcommand{\appname}{{\sc PPatHF}\xspace}
\newcommand{\appnamebold}{{\sc \textbf{PPatHF}}\xspace}

\setcopyright{acmlicensed}
\acmDOI{10.1145/3650212.3652134}
\acmYear{2024}
\copyrightyear{2024}
\acmSubmissionID{issta24main-p755-p}
\acmISBN{979-8-4007-0612-7/24/09}
\acmConference[ISSTA '24]{Proceedings of the 33rd ACM SIGSOFT International Symposium on Software Testing and Analysis}{September 16--20, 2024}{Vienna, Austria}
\acmBooktitle{Proceedings of the 33rd ACM SIGSOFT International Symposium on Software Testing and Analysis (ISSTA '24), September 16--20, 2024, Vienna, Austria}
\received{16-DEC-2023}
\received[accepted]{2024-03-02}

\begin{document}
\begin{sloppypar}

\title{Automating Zero-Shot Patch Porting for Hard Forks}

\author{Shengyi Pan}
\affiliation{%
\institution{The State Key Laboratory of Blockchain and Data Security, Zhejiang University}
\streetaddress{38 Zheda Rd}
\city{Hangzhou}
\country{China}}
\email{shengyi.pan@zju.edu.cn}

\author{You Wang}
\affiliation{%
\institution{Zhejiang University}
\city{Hangzhou}
\country{China}}
\email{3200102524@zju.edu.cn}

\author{Zhongxin Liu}
\authornote{Corresponding Author}
\affiliation{%
  \institution{The State Key Laboratory of Blockchain and Data Security, Zhejiang University}
  \city{Hangzhou}
\country{China}}
\email{liu_zx@zju.edu.cn}

\author{Xing Hu}
\affiliation{%
  \institution{The State Key Laboratory of Blockchain and Data Security, Zhejiang University} %
  \city{Ningbo}
\country{China}}
\email{xinghu@zju.edu.cn}

\author{Xin Xia}
\affiliation{%
 \institution{Huawei}
 \city{Hangzhou}
 \country{China}}
\email{xin.xia@acm.org}

\author{Shanping Li}
\affiliation{
  \institution{The State Key Laboratory of Blockchain and Data Security, Zhejiang University}
  \city{Hangzhou}
\country{China}}
\email{shan@zju.edu.cn}

\begin{abstract} 
Forking is a typical way of code reuse, which provides a simple way for developers to create a variant software (denoted as hard fork) by copying and modifying an existing codebase.
Despite of the benefits, forking also leads to duplicate efforts in software maintenance.
Developers need to port patches across the hard forks to address similar bugs or implement similar features.
Due to the divergence between the source project and the hard fork, patch porting is complicated,
which requires an adaption regarding different implementations of the same functionality.
In this work, we take the first step to automate patch porting for hard forks under a zero-shot setting.
We first conduct an empirical study of the patches ported from Vim to Neovim over the last ten years to investigate the necessities of patch porting and the potential flaws in the current practice.
We then propose a large language model (LLM) based approach (namely \appname) to automatically port patches for hard forks on a function-wise basis.
Specifically, \appname is composed of a reduction module and a porting module. Given the pre- and post-patch versions of a function from the reference project and the corresponding function from the target project,
the reduction module first slims the input functions by removing code snippets less relevant to the patch. 
Then, the porting module leverages a LLM to apply the patch to the function from the target project.
To better elicit the power of the LLM on patch porting, we design a prompt template to enable efficient in-context learning.
We further propose an instruction-tuning based training task to better guide the LLM to port the patch and inject task-specific knowledge.
We evaluate \appname on 310 Neovim patches ported from Vim.
The experimental results show that \appname outperforms the baselines significantly.
Specifically, \appname can correctly port 131 (42.3\%) patches and automate 57\% of the manual edits required for the developer to port the patch.
\end{abstract}



\maketitle

\input{tex/introduction}
\input{tex/motivating_preliminaries}

\input{tex/overview}
\input{tex/approach}

\input{tex/experiment_setup}

\input{tex/experiment_results}
\input{tex/discussion}
\input{tex/relatedwork}
\input{tex/conclusion}

\vspace{-0.1cm}
\section{Data Availability}
The replication package of our work is publicly available at~\cite{replication}.

\vspace{-0.1cm}
\section*{ACKNOWLEDGMENTS}

We thank the anonymous reviewers for their constructive comments to improve the paper.
This research/project is supported by 
the National Natural Science Foundation of China (No. 62202420 and No. 62372398), 
the Ningbo Natural Science Foundation (No. 2023J292),
and the Fundamental Research Funds for the Central Universities (No. 226-2022-00064).
Zhongxin Liu gratefully acknowledges the support of Zhejiang University Education Foundation Qizhen Scholar Foundation.

\balance
\bibliographystyle{ACM-Reference-Format}
\bibliography{reference}

\end{sloppypar}
\end{document}

%% file: tex/introduction.tex
\section{Introduction} \label{sec:introduction}
\vspace{-0.1 cm}

Forking~\cite{zhou2020thesis, fenske2017variant} (a.k.a. clone-and-own) is widely used in both open-source~\cite{ernst2010code, ishio2017source} and industry~\cite{dubinsky2013exploratory} development,
which provides a straightforward and low-cost way for developers to reuse an existing project and tailor it to their own requirements.
In recent years, the rise of social-coding platforms (e.g., GitHub) actively enhances and promotes the ability to reuse existing software via fork-based development paradigm~\cite{zhou2020has, businge2022reuse, businge2022variant}.
Specifically, we follow Zhou~\emph{et al.}~\cite{zhou2020has, zhou2019fork, zhou2020thesis} to refer to the traditional notion of reusing an existing codebase and splitting off an independent variant for new development directions as \emph{hard forks}.
This stands in contrast to a new form of forks (namely \emph{social forks}) that are for isolated development with the intention of contributing back to the mainline~\cite{gousios2014exploratory, gousios2016work}.

Despite the perceived beneﬁts, such as the simplicity and flexibility in creating customized variants of existing softwares,
forking also inevitably leads to
duplicate efforts in software development and maintenance~\cite{ray2012case, robles2012comprehensive, businge2022reuse}.
Specifically, since the design logic and possibly a large portion of code is reused due to the initial copying, 
similar bug fixes and feature implications need to be performed redundantly across a family of hard forks.
Moreover, propagating changes across the hard forks is usually not straightforward
and an error or delayed propagation can possibly compromise the software quality.
Although the source project and the hard forks shares a common codebase initially, their implementations typically diverge over time as the hard forks are intended to be an independent project with a different development direction~\cite{zhou2020has, businge2022variant}.
Thus, when a patch is committed to the source project (or the hard fork), it is not directly applicable (i.e., through a copy-and-paste process) to the hard fork (or the source project).
Instead, porting patches across the family of hard forks requires an adaption regarding different implementations of the same protocol or functionality,
which typically involves identifying the correct location and adjusting the patch according to differences in implementations.
Currently, patches are ported across hard forks manually by the developer on a case-by-case basis, which is time-consuming and often error-prone~\cite{ray2012case, ray2013detecting}.
Several recent empirical works~\cite{zhou2020has, businge2022reuse} point out 
the necessities for developing tools to facilitate coordination of changes across hard forks,
as the existing tools (e.g., \verb|git apply|) 
fail to handle the significant divergence of hard forks.
Moreover, the manual porting process typically faces a long delay~\cite{ray2012case},
which can be serious when the patch is related to security.
Specifically, once a software vulnerability (SV) is patched and publicly disclosed for one of the projects among the hard fork family, malicious actors can easily spot and turn the SV into zero-day attacks against other un-patched projects within the family.
Reid~\emph{et al.}~\cite{reid2022extent} 
recently first investigate the
security risks caused by the code reuse through direct file-level duplication.
They report that \emph{orphan} SVs (i.e., known SVs that have been fixed, but copies still remain vulnerable) are widespread in the OSS ecosystem and often take a long time to be fixed.
Our preliminary study (see Section~\ref{subsec:preliminary}) demonstrates the widespread existence of such security risks within hard forks.

In this work, we aim to fill the gap by automating the patch porting for hard forks.
In practice, it is non-trivial and usually time-consuming to collect a significant number of historically ported patches between the source project and the hard fork.
The ported patches may lack clear documentations in commit messages or being buried in a large commit.
Thus, to ensure the generalizability and practicality of the proposed approach, we focus on a zero-shot setting, i.e., without knowing any historically ported patches.

To understand the necessities of patch porting and the potential flaws in the existing practice, we first conduct an empirical study on Vim~\cite{vim_repo} and Neovim~\cite{neovim_repo}. Vim is a widely-used text editor and Neovim is one of Vim’s most popular hard forks with 71.5K stars on GitHub.
We find that
\ding{182}~Neovim has been porting patches from Vim for nearly 10 years since its creation and the porting rate does not decrease over time.
Over 45\% and 75\% of the total and security patches committed in Vim have been ported into Neovim, respectively.
SVs are only disclosed and recorded for Vim in public SV databases (e.g., NVD), which forces users of Neovim to actively and promptly port the security patches.
\ding{183}~In current practice, the porting is manually done by developers periodically. 
Large porting delays are common, even for high-severity SVs.
For example, over 50\% of the ported patches are delayed for over 180 days.
The porting delay of SV fixes brings serious security risks to Neovim.

It is challenging to automate the patch porting in hard forks, especially under a zero-shot setting. Specifically, we observe two key challenges:
\ding{182}~
The implementations of hard forks can diverge significantly from the source project after a long time evolution (see Section~\ref{subsec:motivating_example}). 
Without a correct understanding of the patch semantics, it is impossible to port the patch with both implementation differences considered and the modification logic maintained.
\ding{183}~Under the zero-shot setting, only the information of the patch from the source project is available and no extra information, such as historically ported patch pairs and relevant test cases, is provided.
The superiority of recent rising large language models (LLM) in understanding the code semantics~\cite{chen2021evaluating, xia2023automated}, as well as its strong generalizability under the zero-shot setting, makes it an ideal solution to tackle the aforementioned two challenges.

Thus, we propose a LLM-based approach (namely~\appname) to automatically \textbf{p}ort \textbf{pat}ches for \textbf{h}ard \textbf{f}orks.
Given a patch in the reference project (typically the source project), 
we port the patch to the target project (typically the hard fork) on a function-wise basis.
\appname contains two modules, i.e., the reduction module and the porting module.
The reduction module first slims the input functions by removing contexts that are less relevant to the patch, enabling \appname to port more patches under the length limit of the LLM.
Then, the porting module leverages a LLM to adapt the patch to the hard fork. 
The LLM can extract common patch logic beyond implementation differences and automate porting at the semantic level.
To effectively instruct the LLM to port patches, 
we design a prompt template to elicit the power of the LLM on the patch porting task by enabling efficient in-context learning (ICL)~\cite{zhao2023survey}.
Moreover, we design an instruction-tuning~\cite{wei2021finetuned} based training
task to further guide the LLM to port the patch and inject project-specific knowledge.
To evaluate our approach, we extract 310 pairs of ported patches from Vim to Neovim after June, 2022 (i.e., the date when the corpus used to pretrain the LLM is collected).
We adopt metrics including accuracy and two metrics from the relevant work~\cite{liu2021just} that measures the edits required to manually port the patch, i.e., Average Edit Distance (AED) and Relative Edit Distance (RED).
\appname outperforms the best performing baselines by a large margin in all metrics. 
Specifically, 42.3\% patches ported by \appname are syntactically equivalent to the the developer ported one, and \appname can reduces 57\% edits on average required by the developer to manually port the patch. 
We also conduct a case study to investigate the effectiveness of \appname in porting security patches.
Moreover, we evaluate the porting performance of \appname on another two hard fork pairs to verify its generalizability.

\vspace{-1 mm}
\begin{itemize}[leftmargin=*]
    \item We propose to automate the patch porting in hard forks. 
    We conduct an empirical study of patches ported from Vim to Neovim over the last ten years, unveiling the necessities and challenges in automating the patch porting across hard forks.

    \item 
    We propose, for the first time, a LLM based approach (namely \appname) to automate patch porting across hard forks under the zero-shot setting.

    \item The experiment results show that \appname outperforms baselines significantly, and can largely reduce the manual efforts required from the developer to port the patch.%

    \item We open source our replication package~\cite{replication} for follow-up works.%
    
\end{itemize} 

%% file: tex/motivating_preliminaries.tex
\section{Motivation and Preliminaries} 
\label{sec:empirical}

In this section, we first provide motivating examples to show the 
usage scenario and challenges of automating patch porting for hard forks.
Then, we conduct a preliminary study to comprehensively investigate the necessities of automating such process.

\input{tex/motivating_example_new}

\input{tex/preliminary_study}

%% file: tex/motivating_example_new.tex
\begin{figure}
    \centering
    \includegraphics[width=0.9\linewidth]{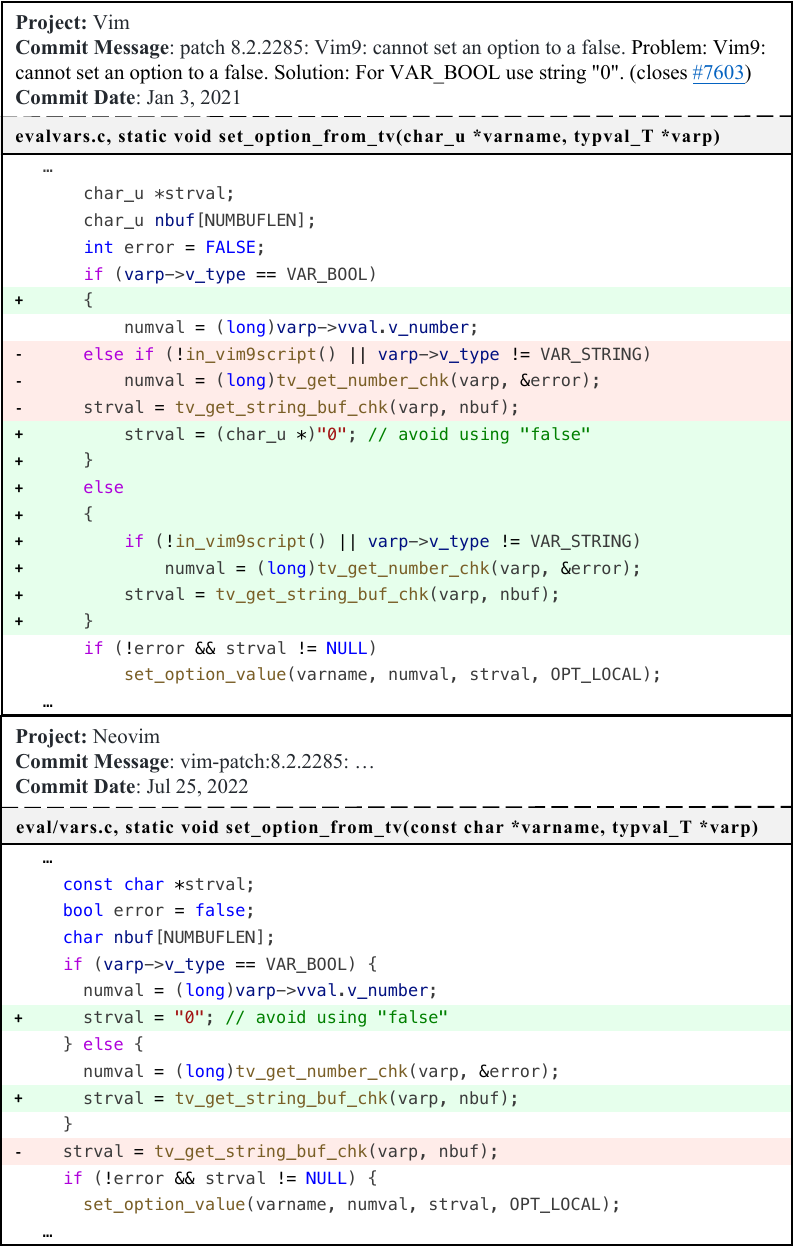}
    \vspace{-0.3 cm}
    \caption{An example of patch porting in Vim-Neovim.}
    \label{fig:motivating}
    \vspace{-0.3 cm}
\end{figure}

\subsection{Motivating Example} \label{subsec:motivating_example}

Figure~\ref{fig:motivating}(a) presents a motivating example collected from Vim (i.e., source project) and Neovim (i.e., hard fork) to demonstrate the patch porting across hard forks.
This patch fixes a bug~\cite{vim_7603} in function \verb|set_option_from_tv()|, which is used to set the option \verb|varname| to the value of \verb|varp| for the current buffer/window.
In the pre-patch version, when \verb|varp| stores the \verb|bool| value False, \verb|numval| will be set to 0 and \verb|strval| will be set to \verb|"false"| after calling the function \verb|tv_get_string_buf_chk()|. 
However, when \verb|numval| is 0 and \verb|strval| contains non-zero characters, \verb|set_option_value()| will raise an error. 
Thus, the patch modifies the logic to set \verb|strval| to \verb|"0"| when \verb|varp| is a \verb|bool| type variable.
This bug also exists in Neovim due to the initial copying of the hard fork.
Therefore, it is necessary to port this patch to Neovim in time.

However, porting this patch is complicated and cannot be achieved by existing tools like \verb|git-apply|~\cite{git_apply}.
Because both the changed lines and the context of the Neovim patch are different from that of the Vim patch.
For example, the function in Neovim does not have the additional \verb|if| restrictions for getting \verb|numval| of non-\verb|bool|-type variables.
In addition, the type of \verb|strval| is \verb|const char*| in Neovim instead of \verb|char_u*| in Vim, making the explicit type casting, i.e., \verb|(char_u *)|, in the Vim patch unnecessary.
In the current practice, Neovim developers manually port the patches on a case-by-case basis, which is time-consuming and error-prone, and typically faces a long delay.
In this example, this patch was first committed in Vim on Jan. 3, 2021 to solve the bug, and was not ported to Neovim until Jul. 25, 2022, delaying for 567.8 days.
Therefore, we argue that it is important and valuable to automate the patch porting for hard forks.

In this work, due to the concern of practicability, we aim to tackle this problem under a zero-shot setting.
This specific task faces the following two key challenges:

    \noindent \textbf{Understand the patch semantics to manage the convergence of logic similarities and implementation differences.}
    Though hard forks share similar design logic with the source project, the detailed implementations diverge over time~\cite{zhou2020has}.
    This means an automated solution for this task needs to correctly understand the patch semantics,
    in order to:
    \ding{182}~Match similar logic between the given code snippets from the source project and the hard fork to locate the code entities to be changed despite of the implementation differences.
    For the example in Figure~\ref{fig:motivating},
    to locate the code entities to be changed in Neovim, an approach needs to tolerate the differences between the \verb|if-else if| 
    block and the \verb|if-else| block, and successfully match them.
    \ding{183}~Adapt the patch according to the implementation differences in the context while maintaining the patch semantics.
    Apart from the relatively straightforward name-space changes (e.g., variable name, variable type, function name), there are more complex modifications involving code structure and logic, e.g., removing code irrelevant to or adding code specific to the context of the hard fork.
    For the example in Figure~\ref{fig:motivating}, the explicit type cast in the added statement (i.e., \verb|strval = (char_u *)"0"|) should be removed when being ported to Neovim. The rational for such change should traceback to the declaration of \verb|strval|, where the variable type is \verb|const char *| instead of \verb|char_u *|.
    {\rm FixMorph}~\cite{shariffdeen2021automated}, the state-of-the-art approach of patch backporting (which is the most relevant task to ours), abstracts the patch into transformation rules at the syntax level.
    It matches the patch and the context based on the same syntactic structure and only allows certain generalizations of the identifier names.
    It can not handle the patch porting for hard forks where the syntactic structure changes greatly.

    \noindent \textbf{Automate the patch porting given only the information of the patch from the source project, without additional prerequisite of knowing any extra information, such as historically ported patches, available test cases.}
    As mentioned in the introduction, the extra information is not always available or expensive to collect in practice. 
    This renders most existing approaches in the relevant works invalid. 
    For example,
    automated program transformation methods 
    including GetaFix~\cite{bader2019getafix} and Phoenix~\cite{bavishi2019phoenix} 
    require to learn the transformation rules and determine the abstraction level from multiple similar existing patches.
    PatchWeave~\cite{shariffdeen2020automated} transforms the patch by concolic executions and requires available test cases.

We observe unique opportunities in leveraging a LLM to solve the aforementioned challenges:
\ding{182}
Endowed with the ability of capturing the semantic information of the given code snippet~\cite{li2023starcoder, xu2022systematic, chen2021evaluating},
the LLM is able to capture the implementation differences and automate the patch porting based on similar high-level code semantics.
\ding{183}
The \emph{off-the-shelf} LLM has already been proven to significantly outperform the existing approaches on various code generation tasks~\cite{fan2023automated, xia2023automated, chen2021evaluating}.
The generalizability of the LLM, especially the strong performance under a zero-shot setting~\cite{larochelle2008zero}, 
makes the LLM an ideal solution to automate patch porting for hard forks.
We leverage the LLM to build our approach, namely \appname, an effective and generalizable solution to automate patch porting.
Note that \appname correctly ports the patch demonstrated in the example.

%% file: tex/preliminary_study.tex
\vspace{-0.2cm}
\subsection{Preliminary Study}
\label{subsec:preliminary}
\vspace{-0.1cm}
We conduct a preliminary study on the patches ported from Vim to Neovim to understand the current practice of patch porting.

\begin{figure}
    \centering
    \includegraphics[width=0.8\linewidth]{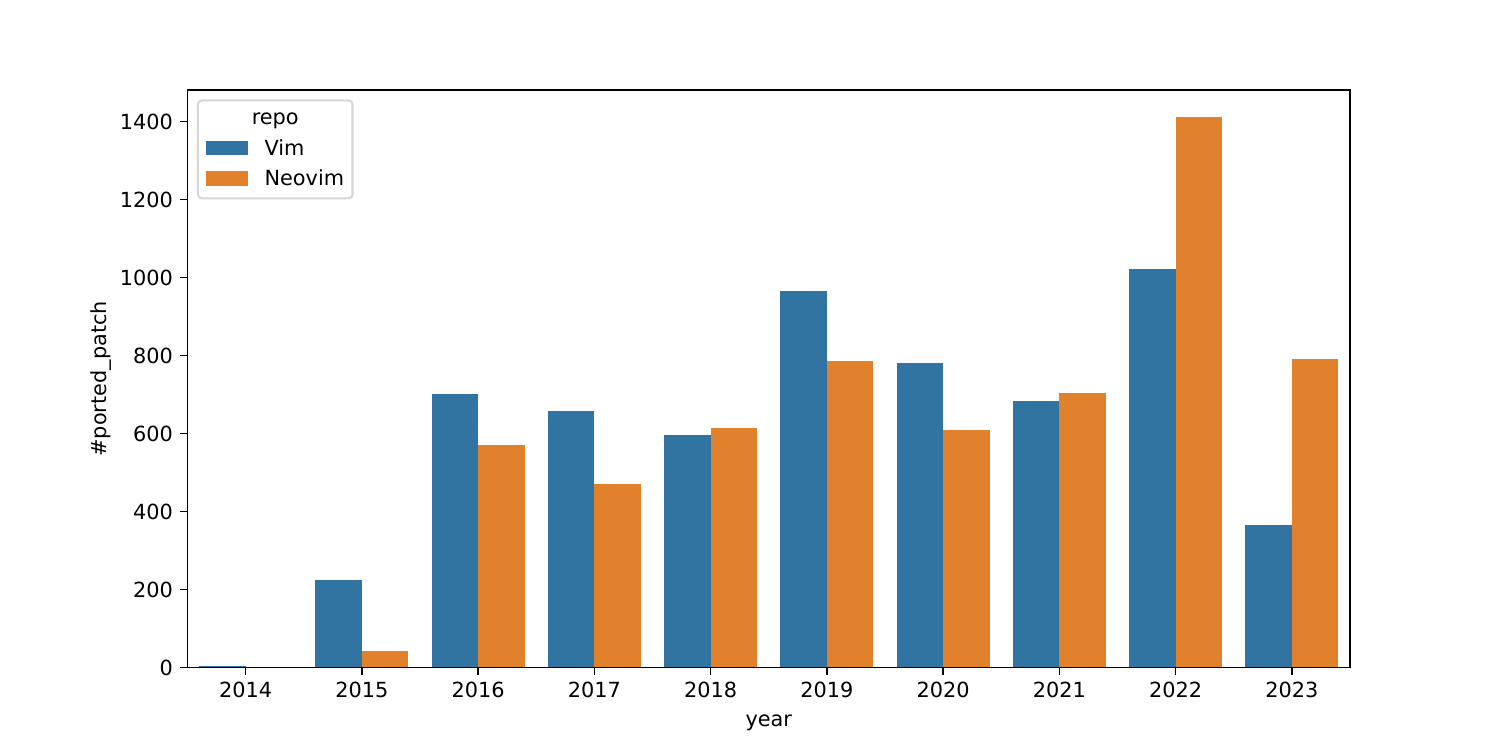}
    \vspace{-0.3 cm}
    \caption{Distribution of ported patches over the years}
    \label{fig:ported_patch_distri_over_year}
    \vspace{-0.5 cm}
\end{figure}

\noindent \textbf{Porting patches from source project is necessary for hard forks.}
We identify 6,487 Neovim patches ported from Vim, spanning from April. 2, 2014 to Aug. 24, 2023.
Figure~\ref{fig:ported_patch_distri_over_year} presents the distribution of ported patches over the years.
The Neovim repository was created on GitHub on Jan. 31, 2014~\cite{neovim_repo}. Two month after its creation, Neovim began porting patches from Vim (i.e., its source project), and keeps porting until the time we collect our dataset (i.e., Sept. 1, 2023),
and the number of ported patches does not decrease over time.
Besides, in each year before 2022, the number of patches ported to Neovim has generally been lower than the required one, indicating a considerable delay in patch porting and there may still exist historical Vim patches that need to be ported.
The ported Neovim patches correspond to nearly 46\% of the patches committed to Vim. 
We find that most patches ported to the hard fork relate to the bug fixing caused by the reuse of the code from the source project.
Among the bugs discovered in the source project, one critical case are software vulnerabilities (SVs).
We collect 175 Common Vulnerabilities and Exposures (CVE) records related to Vim after year 2014 (i.e., when the Neovim project is initiated) according to the Ubuntu security advisory~\cite{ubuntu_security_advisory_vim}.
By cross-referencing CVEs with ported patches, we find that over 76\% of security patches have been ported to Neovim.
Moreover, we find that these SVs are only disclosed for Vim (i.e., the source project) but not Neovim. 
This aligns with the observations made by Wang \emph{et al.}~\cite{wang2019detecting} that although similar SVs may exist in software with similar functionalities (e.g., SSL libraries) due to similar design logic or code clone, typically only one CVE is disclosed for one of the particular software.
Such practices make the threats to hard forks unlikely to be alerted timely or even disregarded~\cite{wang2019detecting}, e.g., the Software Component Analysis (SCA) tools scan the codebase for known vulnerable components by referring SV databases (e.g., NVD) ~\cite{chen2020machine, wu2023understanding}.
To reduce the exposure to zero-day attacks~\cite{bilge2012before}, it requires the stakeholders of hard forks to actively monitor the SVs disclosed for upstream projects and promptly port the corresponding patches.

\begin{figure}
    \centering
    \includegraphics[width=0.8\linewidth]{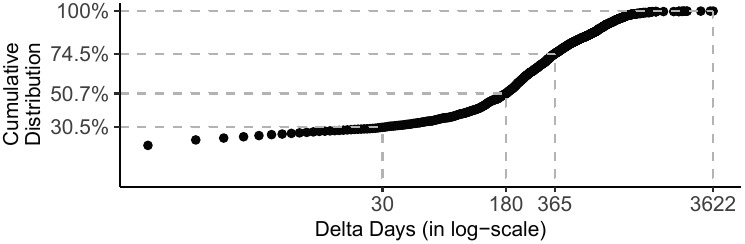}
    \vspace{-0.3 cm}
    \caption{ECDF plot of delta days of patch porting}
    \label{fig:delta_days_patch_porting}
    \vspace{-0.5 cm}
\end{figure}

\noindent \textbf{It takes a large delay for hard forks to port the patches in current practices.}
We further investigate the delay between the time a patch is committed in Vim (i.e., the source project) and it being ported in Neovim (i.e., the hard fork). 
Figure~\ref{fig:delta_days_patch_porting} presents the cumulative distribution function (ECDF) of delta days before porting patches.
Nearly 50\% of patches took more than 180 days to be ported into Neovim, and over 25\% even took more than a year.
This indicates that a large portion of issues discovered and fixed in Vim will still persist in Neovim for an extended long period of time.

When it comes to SVs, such delay in patch porting could be fatal. The delay leaves an exploit window wide open for malicious actors to launch 0-day attacks against Neovim by simply referring to disclosed Vim SVs.
We also examine the security patches that have been ported to Neovim (134 in total). The median porting delay is 37 days and over 25\% of security patches take more than three months to be ported.
Although Neovim ports security patches in a more urgent manner compared to ordinary defects, we contend that such delay is still fatal, considering that information about the SV in Vim has already been publicly disclosed (possibly including a functional exploit).
Taking CVE-2022-0413~\cite{CVE-2022-0413} as an example, it is a \emph{Use After Free} vulnerability in Vim with a \emph{HIGH} severity.
The vulnerability was publicly disclosed on NVD on Jan. 30, 2022. 
However, it was left in the Neovim codebase for another 64 days before the patch porting~\cite{CVE-2022-0413_neovim_patch}.
Once publicly disclosed, this SV becomes a hot spot for malicious actors~\cite{elbaz2020automated}, especially given its high severity.
Moreover, the SV was disclosued with links of the Vim patch~\cite{CVE-2022-0413_vim_patch} and a verified proof-of-concept (PoC)~\cite{CVE-2022-0413_POC} listed as external references.
Such information further ease the difficulties for malicious actors to turn the SV into a 0-day attack against Neovim.
Over 70\% percent of the security patches are ported into Neovim after the disclosure of the Vim SV, and nearly 85\% of SVs are rated with \emph{HIGH} or \emph{CRITICAL} severity.
By automating the patch porting for hard forks, we can significantly reduce the manual effort required in the porting process, thus alleviating the delay in current practices.

%% file: tex/overview.tex
\vspace{-0.1cm}
\section{Problem Formulation} \label{sec:overview}

This work aims to automatically port the patch across the family of hard forks (typically from the source project to the hard fork).
Specifically, 
we conduct patch porting at the function granularity,
as functions represent self-contained, fine-grained blocks of code.
Specifically, given the pre-patch and post-patch versions of a modified function from the source project, namely $ (f_{s},f^{'}_{s} )$, and the pre-patch version of the corresponding function in the hard fork, namely $f_{f}$, the LLM is utilized to automatically \emph{port} the patch derived from differentiating $f_{s}$ and $f^{'}_{s}$ and apply it to $f_{f}$:

\vspace{-1mm}
\begin{equation}
    port\left ( f_{s},f^{'}_{s}, f_{f}\right ) = f^{'}_{f}
    \label{eq:classification_prob}
\end{equation}
\vspace{-1mm}

\noindent where $f^{'}_{f}$ is the expected post-patch version of the function in the hard fork.
Since we aim to port the patch from the source project to facilitate a similar code change (e.g., bug fixing, feature implication) in the hard fork, it is essential for $port$ to maintain the functionality of the patch while correctly adapting it based on the context of the hard fork.
Besides, we focus on a zero-shot scenario, i.e., automating the patch porting without knowing any historically ported patch of the given pair of source project and hard fork.

%% file: tex/approach.tex
\vspace{-0.1cm}
\section{Approach} \label{sec:approach}
\vspace{-0.1cm}

\begin{figure}
    \centering
    \includegraphics[width=0.9\linewidth]{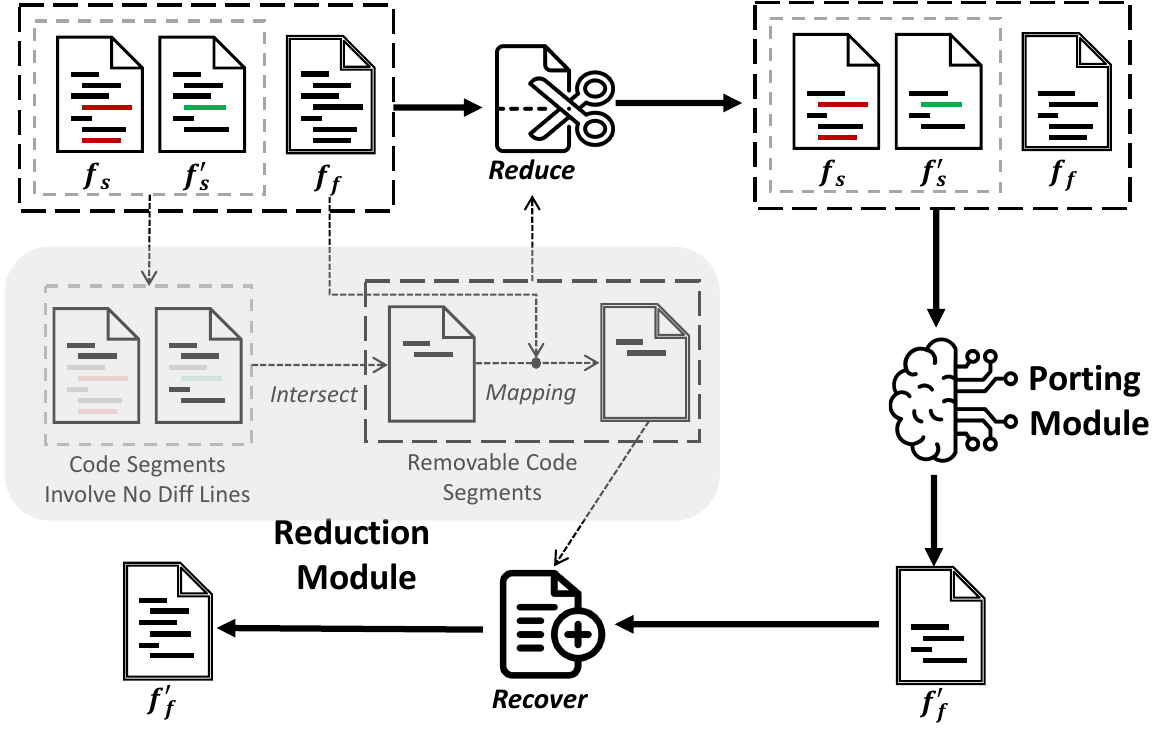}
    \vspace{-0.5 cm}
    \caption{Overview of \appnamebold}
    \label{fig:approach_overview}
    \vspace{-0.5 cm}
\end{figure}

In this section, we introduce our proposed approach, namely \appname.
Figure~\ref{fig:approach_overview} presents the overview of \appname, which is composed of a reduction module and a porting module. 
Specifically, the reduction module first slims the input functions by removing code snippets that are less relevant to the patch.
Then, the porting module adapt the patch derived from the pair of pre and post patch functions from the source project to the hard fork by generating the modified function.
Finally, the generated post-patch function is recovered by reintegrating the code snippets that are previously removed by the reduction module. 

\input{tex/slicing}

\input{tex/llm}

%% file: tex/slicing.tex
\vspace{-0.1cm}
\subsection{Reduction Module} \label{subsec:reduction}
\vspace{-0.1cm}

Our reduction module discards code segments in the functions that are less relevant to the patch, 
allowing the LLM to port more patches under its length limit.
The modified functions are generally very lengthy. 
We show in our experiments that under the length limit of 2k, the LLM is only able to port 26.8\% of the patches.
Moreover, the reduction module can largely reduce the computational costs and allowing the LLM to focus on the key code related to the modification.

The upper left part of Figure~\ref{fig:approach_overview} illustrates the framework of our reduction module.
Briefly speaking, we first extract code segments in the function from the source project that do not overlap with code lines modified by the patch. 
We refer to these code segments as removable segments.
Subsequently, we map these removable segments from the source project to the hard fork. 
Using the acquired the removable segments for functions in both the source project and the hard fork, we slim the input functions $(f_s, f^{'}_{s}, f_{f})$ by removing these segments and later integrate them back to recover the $f^{'}_{f}$ generated by the porting module.

\noindent \textbf{Extracting Removable Segments. }
We start by extracting removable segments for the function from the source project, represented in the form of a list of Abstract Syntax Tree (AST) nodes.
Arbitrarily reducing nodes in an AST may result in evident syntax errors (e.g., reducing an identifier in an expression statement) or distortion of the original function (e.g., reducing an if condition while preserving the if body), potentially confusing the LLM and leading to wrong porting results.
Thus, during the extraction of removable segments, we only focus on sub-trees whose root nodes are of compound statement types (e.g., if, for, and while statement). 
These sub-trees within the AST constitute structurally complete units, ensuring that their reduction does not distort the original function.
Moreover, code segments corresponding to the sub-trees of compound statement are generally longer than those of simple statement, alleviating the potential errors when mapping these segments from the source project to the hard fork.

Specifically, we first conduct a diff operation between $f_s$ and $f^{'}_{s}$ to obtain the statements deleted and added by the patch, denoted as $stm_{del}$ and $stm_{add}$, respectively.
Next, we traverse the AST of $f_s$ to create a list of mutually exclusive sub-trees that are not overlapped with $stm_{del}$.
For $f^{'}_{s}$ and $stm_{add}$, we follow the same procedure to obtain another list. 
Finally, we calculate the intersection of these two lists to obtain a list of sub-trees within the AST, which represents all code segments that can be removed in both $f_s$ and $f^{'}_{s}$ while retaining the modified statements.
Such matching between the two lists is straightforward because the pre- and post-patch function are identical after excluding the modified statements.

\noindent \textbf{Mapping Removable Segments. }
In this part, we map the removable segments of $f_{s}$ to $f_{f}$.
These two pre-patch functions are semantically similar, but usually with different implementation details.
Thus, we perform the mapping based on text similarity.
Specifically, to determine the similarity between two code segments from $f_s$ and $f_f$ (denoted as $s_s$ and $s_f$), we first normalize them (e.g., removing empty characters, lowering the case of alphabetic characters) and then calculate the similarity score using the following formula.

\vspace{-1mm}
\begin{equation}
    similarity(s_s, s_f)=\frac{maxlen(s_s, s_f) - edit\_distance(s_s, s_f)}{maxlen(s_s, s_f)}
    \label{eq:similarity}
\end{equation}
\vspace{-1mm}

\noindent where $maxlen$ is length of the longer code segment, $edit\_distance$ computes the Levenshtein distance between the two code segments. 

Given a removable segment in $f_s$ as the target, to find its matching code segment among the candidate segments in $f_f$, 
we initially calculate the similarity score between the target and each candidate, and eliminate candidates with a similarity score lower than a specified threshold $thres_{self}$. 
Instead of solely relying on the local information (i.e., the code segment itself) to conduct matching, we further include the context information to eliminate the possible matching errors, 
i.e., considering the similarity between code snippets corresponding to the parent AST node of the segments to be matched (denoted as the parent code snippet).
Specifically, since similar code segments may appear in multiple different locations within the function, we also take into considerations of the relative position of the code segment within the its parent code snippet. 
We first divide the parent code snippets into two halves (i.e., the preceding context and the following context) based on the position of the code segments to be matched.
Then, we calculate the similarity scores between the two preceding halves and the two following halves, respectively.
Finally, we average the two scores
weighted by the length of the corresponding half and use it as the final similarity score between the parent code snippets. 
We further eliminate candidates whose parent similarity is under the specified threshold $thres_{parent}$.
If there are still candidates left, we consider the one with the highest parent similarity as the match.

For each removable segment of $f_s$ acquired in the extraction phase, 
we attempt to retrieve the corresponding segment in $f_f$, using the aforementioned matching process. 
If a match is retrieved, we pair it with the removable segments of the function from the source project to create a tuple. 
This results in a list of tuples of removable segments, which is used to reduce the input functions.

\noindent \textbf{Reducing \& Recovering.}
We first index the removable tuples acquired in the mapping process, 
and then simultaneously replace these segments with placeholder comments in the form of \verb|/* Placeholder_{index} */| for each of the three input functions $(f_s,f^{'}_{s},f_f)$. 
After the porting module generates the patched function $f^{'}_{f}$, we reintegrate these previously removed code segments back into their due positions to recover the generation. 
Specifically, for each placeholder comment found in the generated $f^{'}_{f}$, we utilize the parsed index to retrieve the corresponding removed code segment, and use it to replace the comment.

%% file: tex/llm.tex
\vspace{-0.2cm}
\subsection{Porting Module} \label{subsec:llm_module}
\vspace{-0.1cm}

\begin{figure}
    \centering
    \includegraphics[width=\linewidth]{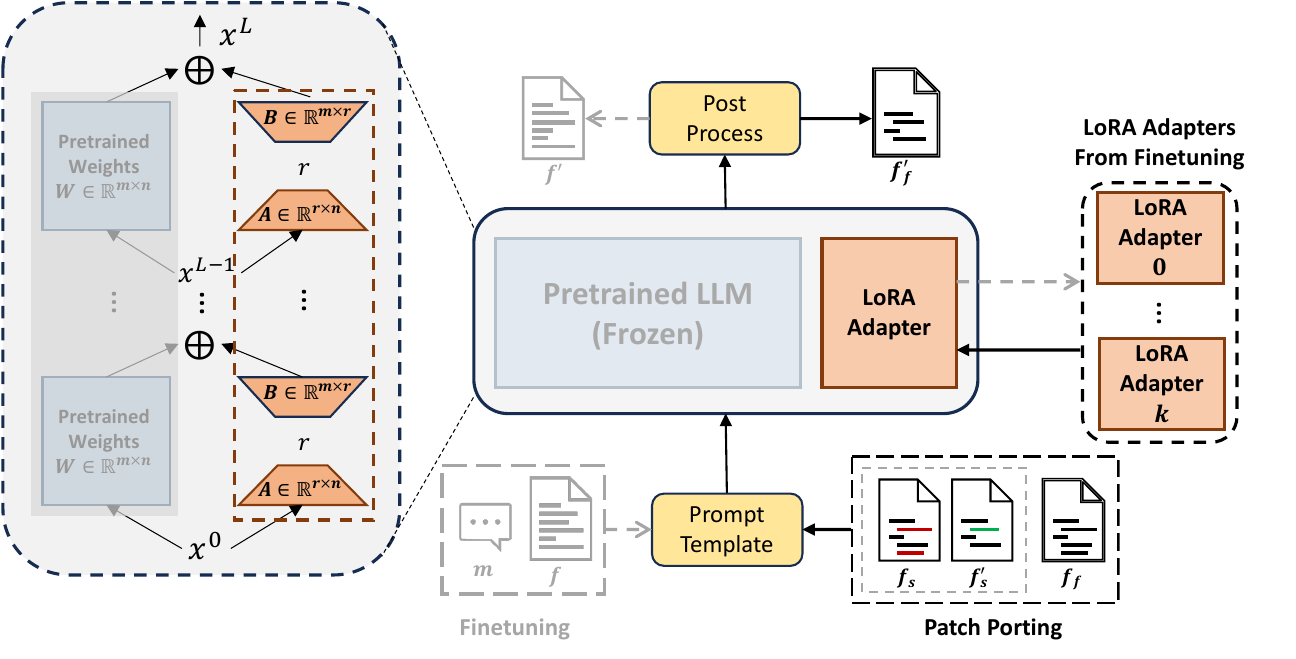}
    \vspace{-0.4 cm}
    \caption{Framework of the Porting Module}
    \label{fig:porting_module}
    \vspace{-0.3 cm}
\end{figure}

\begin{figure}
    \centering
    \includegraphics[width=0.8\linewidth]{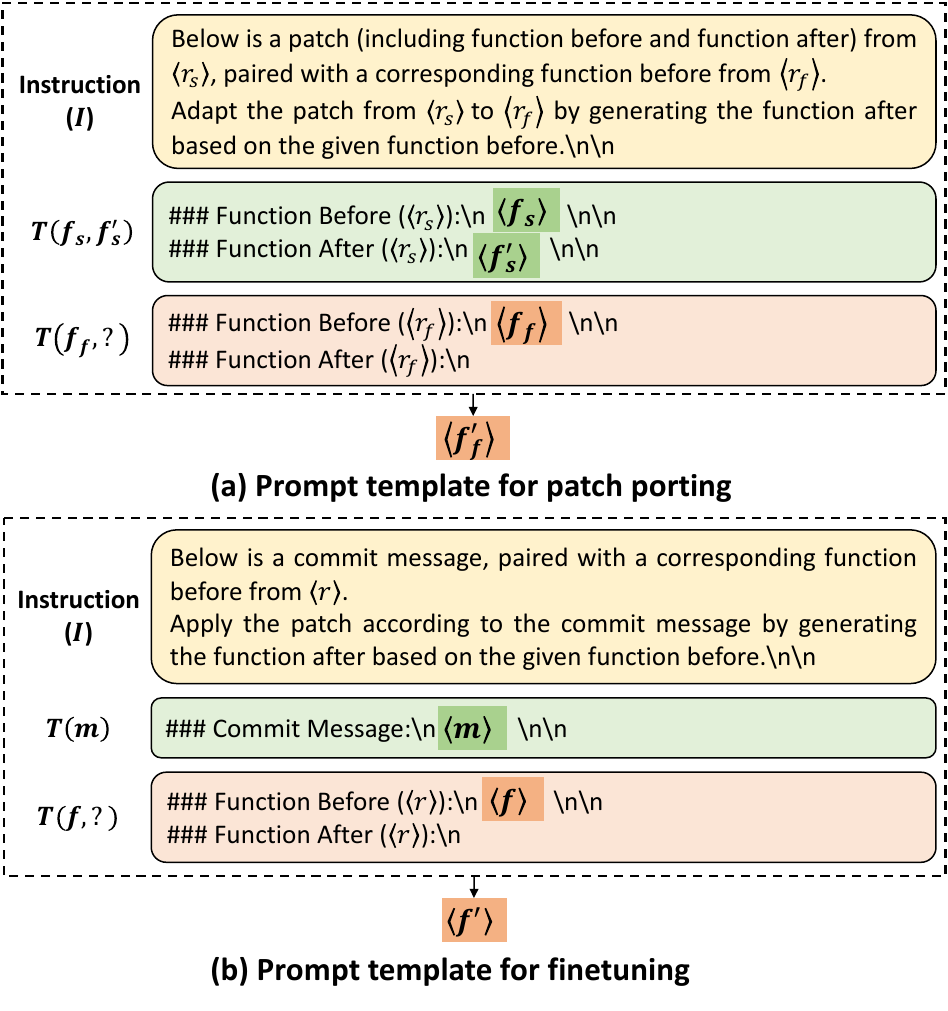}
    \vspace{-0.5 cm}
    \caption{Designed prompt template for the patch porting and the finetuning task. $\mathbf{\left \langle \cdot \right \rangle } $ stands for a placeholder.}
    \label{fig:prompt_template_merged}
    \vspace{-0.5 cm}
\end{figure}

After pretraining on large-scale corpora, LLMs are endowed with abilities as general and capable task solvers~\cite{zhao2023survey}.
To better elicit LLM's ability on our patch porting task,
we design the prompt template following~\cite{alpaca} (see Figure~\ref{fig:prompt_template_merged}(a)). 
We give a natural language instruction of the task first and followed by the task input, i.e., the pre- and post-patch versions of the function in the source project $( f_{s},f_{s}^{'} )$ and the pre-patch version of the corresponding function in the hard fork $f_{f}$.
Specifically, we instruct the LLM to generate the post-patch version of the function in the hard fork $f_{f}^{'}$ by extracting the patch logic from $( f_{s},f_{s}^{'} )$ and apply it to $f_{f}$.
Moreover, the format of our designed prompt template aligns with the guidance for in-context learning (ICL)~\cite{brown2020language, zhao2023survey}:

\vspace{-0.1cm}
\begin{equation}
    \mathrm {LLM} ( I, \underset{demo }{\underbrace{T ( f_{s}, f^{'}_{s} )}} , \underset{query}{\underbrace{ T (f_{f}, ?)}}) \to f^{'}_{f} 
    \label{eq:ICL_template}
\end{equation}
\vspace{-0.1cm}

\noindent where $I$ is the task instruction and $T(\cdot)$ is the template used to format the input into the prompt.
In some point, the pre- and post-patch versions of function from the source project $( f_{s},f_{s}^{'} )$ can be regarded as a demonstration, which shows the LLM how to patch the $f_{f}$.
The output is left as a blank to be completed by the LLM through imitating the modifications presented by the demonstrations from the source project.
Thus, the designed prompt template enables efficient ICL, which helps to better adapt the general ability of the LLM acquired from pretraining to patch porting.

Although with proper prompt engineering, the LLM already demonstrates powerful capabilities in various tasks~\cite{brown2020language, sanh2021multitask} without gradient update,
the prompt engineering faces the following limitations~\cite{zhao2023survey}:
1) limited window size for context;
2) there lacks a clear and deterministic practice regarding prompt design for a new task like patch porting. Even subtle changes in prompt can strongly impact LLM performance~\cite{sclar2023quantifying}.
Thus, 
we further propose an instruction-tuning~\cite{wei2021finetuned} based training
task to
1) better adapt the LLM to the patch porting task;
2) inject project specific knowledge into the LLM.
In our designed finetuning task, 
we finetune the LLM to patch a function (i.e., generate the post-patch version of the given function) based on the given commit message.
The prompt template used to format the instances is shown in Figure~\ref{fig:prompt_template_merged}(b).
The LLM is trained to optimize the sequence-to-sequence loss (see Section~\ref{subsec:experiment_setting} for more training details).
To collect the training instances, we crawl the commit history of both the source project and the hard fork, and further select commits suitable (e.g., with high quality commit message) for the instruction tuning (refer to Section~\ref{subsec:data_preparation_middle} for details of the data preparation).
Note that the design of our finetuning task is different from the typical one that trains the model using the exact same input and output with the evaluation. Our design conforms to the principle of zero-shot setting to avoid compromising the generalizability.
Specifically, we do not require the information of any known ported patch pairs (which are not readily available), but the commit history which can be easily obtained for most modern softwares (e.g., using version control systems like git).
Besides, our proposed finetuning task is closely related to the patch porting, i.e., applying the patch (either derived from the commit message or a pair of pre- and post-patch functions) to a given function to generate its post-patch version. 
The prompt template (both the task instruction and the format) for these two tasks also have a lot in common (see Figure~\ref{fig:prompt_template_merged}).
Thus, by tuning with the proposed training task, we encourage the LLM to better follow the instructions to perform patch porting, alleviating the complexity of prompt engineering~\cite{wei2022emergent}.
Tuning can also help to align the output of the LLM with our expectations, alleviating the known weaknesses including completing the input without addressing task or generating repetitively~\cite{ouyang2022training, chung2022scaling}.
More importantly, apart from better adapting the general ability of the LLM (acquired from pretraining) towards the patch porting task, instruction-tuning is proven to be an effective approach to inject domain-specific information into general LLMs~\cite{wu2023bloomberggpt, liu2023goat}.
The context information at the function scope may not be sufficient to ensure a successful porting (e.g., a changed function call).
A general understanding of both the source project and the hard fork (e.g., implementations, coding styles) is a prerequisite of a correct porting.
Our proposed finetuning task effectively helps the LLM familiar with the projects, thus turning it into a domain-specific expert.

Considering the huge cost of full parameter tuning,
we adopt Low-Rank Adaption (LoRA)~\cite{hu2021lora} as a lightweight solution to facilitate finetuning.
Specifically, considering a general form of the updating parameter matrix $\mathrm {W} \gets \mathrm {W}+  \triangle \mathrm {W}$, LoRA freezes the original matrix $\mathrm {W} \in \mathbb{R}^{m\times n}$ while approximating the update matrix $\mathrm {\triangle W} \in \mathbb{R}^{m\times n}$ by low-rank decomposition, i.e., $\triangle  \mathrm {W} = \mathrm {B}\mathrm{A}$, where $\mathrm {B} \in  \mathbb{R}^{m \times r}, \mathrm {A} \in  \mathbb{R}^{r \times n}$, and the rank $r \ll \min \left ( m, n \right ) $.
Instead of tuning the pretrained model weights ($\mathrm{W}$), LoRA injects and tunes the low-rank decomposition matrices ($\mathrm{A},\mathrm{B}$) instead,
thus significantly reducing the number of trainable parameters (11.4\% compared with full tuning in our implementation) when adapting the LLM to patch porting task (as shown in Figure~\ref{fig:porting_module}).
The reduction in trainable parameters, in turn, significantly lowers the computational cost required for tuning the LLM, making our proposed finetuning task more cost-effective.
Most importantly, the separation between the frozen LLM and LoRA adapters (i.e., the injected low-rank decomposition matrices) ensures efficient deployment against various pairs of porting projects.
Specifically, the pretrained LLM can be shared as a base, allowing for the easy switch between small LoRA adapters to accommodate different projects.

%% file: tex/experiment_setup.tex
\vspace{-0.1cm}
\section{Experiment Setup} \label{sec:experiment_setup}
\vspace{-0.1cm}
In this section, we first introduce our data collection procedure. Then, we describe our experiment settings. 

\input{tex/dataset}
\input{tex/experiment_setting}

%% file: tex/dataset.tex
\vspace{-0.1cm}
\subsection{Data Preparation for Patch Poring} \label{subsec:data_preparation_porting}
\vspace{-0.1cm}

We use Vim-Neovim as a specific example of to investigate the patch porting from the source project to the hard fork.
We select Vim-Neovim based on the following reasons:
\ding{182}~Vim-Neovim is a typical case of hard fork, i.e., successful branching with differentiation~\cite{zhou2020has, robles2012comprehensive}.
Both Vim (the source project) and Neovim (the hard fork) succeed as well-known and widely used text editors (with 33.5k and 71.1k GitHub stars respectively). 
Both projects remain active for a long time (nearly ten years).
\ding{183}~Neovim clearly records the patches that are ported from Vim and follows a one-to-one porting style (i.e., one Vim patch is ported into Neovim as an individual patch).
Since it has been nearly ten years since the initial fork of Neovim, porting patches from Vim has become a routine of the Neovim development and there are abundant historical ported patches.
These characteristics ease the difficulties in collecting and analyzing the ported patches, as well as conducting experiments to evaluate the porting performances.

We build dataset in the form of a quadruples $( f_{s}, f_{s}^{'}, f_{f}, f_{f}^{'} )$.
The first three functions are the task input, and the last one is the expected output (see Section~\ref{sec:overview} for a detailed task formulation).
We build our patch porting dataset based on the following steps:

\noindent \textbf{STEP 1: Collect project history commits.}
We first extract the entire commit history of Vim and Neovim util Sept. 1, 2023.
We use Git to traverse the historical commits and collect the relevant information (e.g., commit hash, message, and the committed date).

\noindent \textbf{STEP 2: Pair the ported Neovim patches with the source one from Vim.}
Based on the manual check of Neovim commits, we observe that the commit message of the patches ported from Vim share a fixed
format, where the message title starts with ``vim-patch:" followed by the same patch id used in Vim and the message body posts the link to the source patch in Vim.
Thus, we filter the collected Neovim commits by matching commit messages with the above two patterns to get the patches that are ported from Vim.
As a result, we obtain 6,487 ported patches in Neovim, and out of which, we are able to retrieve the corresponding valid Vim commit hash for 5,998 patches.
Some commit messages of the ported Neovim patch may post the Vim issue id instead of the commit hash or simply forget to post any reference link.

\noindent \textbf{STEP 3: Filter ported patch pairs to build the test set.}
Since we focus on automating the patch porting with a zero-shot setting, there are no known ported patch pairs available for model training or validation.
We build our test set using the patches ported into Neovim after July 1, 2022.
Becuase, the corpus (namely The Stack~\cite{kocetkov2022stack}) used to pretrain StarCoder~\cite{li2023starcoder} (the LLM used in our implementation) is collected from extensive GitHub repositories (downloaded before July 2022).
Thus, we only include the patches ported after the collection date of the pretrain corpus to avoid possible training-to-test data leakage, which has been shown to compromise the validity of the evaluation~\cite{xu2022systematic, elangovan2021memorization}.
We further restrict our test set into the patches that only modifies a single function in both Vim and Neovim.
Specifically, for each patch, we only focus on the modifications to the source code, excluding changes to comments in *.c files (excluding changes to documentations and header files). 
Note that the single-function restriction is mainly to simplify the evaluation as \appname automates the patch porting at the function granularity (see Section~\ref{sec:overview}).
We further conduct a case study in Section~\ref{subsec:discuss_security} to verify the validity of \appname when generalizing to patches involving multiple function changes.

\noindent \textbf{STEP 4: Preprocess the ported patch pairs.}
For each filtered patch pair, we further extract the functions before and after the patch in Vim and Neovim, respectively.
We preprocess each function by removing the comments and empty lines.
Finally, our test set includes 310 pairs of ported patches.

\vspace{-0.2cm}
\subsection{Data Preparation for Finetuning} \label{subsec:data_preparation_middle}
\vspace{-0.1cm}

In the proposed finetuning task (see Section~\ref{subsec:llm_module}), 
we tune the LLM to patch the given function $f$ to generate its post-patch version $f^{'}$ according to the commit message $m$.
We build the dataset in the form of a triplet $( f,f^{'},m ) $ by collecting project history commits with single-function change.
Note that we only keep the commits before July 1, 2022 for finetuning to make sure there is no leakage of the test data.
Since the commit message is included in the prompt to instruct the LLM to patch the function (see Figure~\ref{fig:prompt_template_merged}(b)),
we clean and format the commit message to only keep the natural language descriptions while removing code snippets, commit hashes, etc.
We further exclude commits with low-quality messages (e.g., cleaned messages containing less than five words) from finetuning.
These messages are unable to provide valid instructions to guide the LLM patch the function, thus harming the effectiveness of finetuning.
Finally, our finetuning dataset contains 2,747 and 2,082 commits from Vim and Neovim, respectively.

%% file: tex/experiment_setting.tex
\vspace{-0.2cm}
\subsection{Experiment Setting}\label{subsec:experiment_setting}
\vspace{-0.1cm}
The experimental environment is a server with NVIDIA A800 GPUs, Intel Xeon 8358P CPUs, running Ubuntu OS.

\noindent \textbf{Implementation Details.} 
We use StarCoder~\cite{li2023starcoder}, a recent and powerful foundation model of code, to implement the porting module of \appname.
We download the public accessible pretrained weights (with 15.5B parameters) from the Hugging Face Transformer library~\cite{starcoder_huggingface} to initialize our model.
In the finetuning, we tune the StarCoder to optimize the sequence-to-sequence loss in a supervised way.
Specifically, the LLM is trained with the causal language modelling (CLM) task~\cite{radford2018improving}, but with the CLM loss only applied to the output sequence (i.e., the completion).
We mask the CLM loss of the input sequence (i.e., the prompt) to ensure the model is only optimized to generate the expected output.
We adopt LoRA to tune the LLM with the rank set to 8 following~\cite{alpaca_lora}.
we use AdamW~\cite{loshchilov2017decoupled} as the optimizer. The learning rate is set to $2e^{-5}$ following~\cite{alpaca}. 
During the training, we adopt a cosine learning rate scheduler with the warmup ratio set to 0.03 following~\cite{alpaca}.
We tune the LLM for 10 epochs during the finetuning.

\noindent \textbf{Baselines.}
We include the following four baselines:
\ding{182}~\emph{Origin} is a special baseline for patch porting, which directly outputs the given pre-patch version of the function in the hard fork without applying any modifications.
It measures the default manual efforts to port the patch.
By comparing \appname with \emph{Origin}, we can know whether the generated post-patch functions are closer to the ground truth than the pre-patch version.
\ding{183}~\emph{git-apply}~\cite{git_apply} is the tool from the Git suit.
\ding{184}~\emph{patch}~\cite{linux_patch_tool} is the tool from GUN Diffutils.
These two tools are used to apply a given patch to the target file.
With these baselines, we examine whether porting patches across hard forks can be done with a simple copy-and-paste process.
\ding{185}~\emph{FixMorph}~\cite{shariffdeen2021automated} is the state-of-the-art approach in patch backporting.
It is the most relevant work to our proposed zero-shot hard fork patch porting task,
which backports a patch from the Linux mainline version to older versions at the syntax level.
\ding{186}~\emph{StarCoder}~\cite{li2023starcoder} is one of the most advanced open-source code LLMs.
To the best of our knowledge, no existing neural approach can be directly applied to our patch porting task, especially under a zero-shot setting.
Since we are first to introduce the task of patch porting in hard forks, we also investigate the use of the \emph{off-the-shelf} LLM for this task by including StarCoder as one baseline.
Moreover, we implement our porting module using the StarCoder as the underlying LLM.
Thus, by comparing \appname with the StarCoder, we can better understand the shortcomings of directly applying the \emph{off-the-shelf} LLMs in solving the patch porting task and the effectiveness of our design.
We use the same prompt (see Figure~\ref{fig:prompt_template_merged}(a)) for StarCoder and \appname.

\noindent \textbf{Evaluation Metrics.}
To automate the patch porting in hard forks, \appname aims to generate the post-patch version of the given function in the hard fork based on the patch from the source project.
We evaluate the performance by adopting the metrics, i.e., Accuracy, AED, and RED, used in the relevant works regarding
1) patch backporting for Linux kernel~\cite{shariffdeen2021automated},
2) comment updating~\cite{liu2021just}, which shares a similar task setting with ours, i.e., updating a given code comment based on the patch.
Accuracy is used to present to what extent an approach can port the patch \emph{correctly}.
We refer a \emph{correct} porting to the generated post-patch function $\hat{f}_{f}^{'}$ that is syntactically equivalent to the developer patched one $f_{f}^{'}$.
Apart from the strict identical match, Average Edit Distance (AED) and Relative Edit Distance (RED) measures the edits developer still need to perform to correctly patch the function after using a patch porting tool, which provides a more detailed evaluation of the porting performance. 
Specifically, AED and RED are calculated as follows:

\vspace{-1mm}
\begin{equation}
\begin{aligned}
AED &= \frac{1}{N}\sum_{i = 1}^{N} edit\_distance\left ( \hat{f}_{f}^{'(i)}, f_{f}^{'(i)} \right ) \\
RED &= \frac{1}{N}\sum_{i = 1}^{N} \frac{edit\_distance\left ( \hat{f}_{f}^{'(i)}, f_{f}^{'(i)} \right )}{edit\_distance\left ( f_{f}^{(i)}, f_{f}^{'(i)} \right )} 
\end{aligned}
    \label{eq:metrics}
\end{equation}
\vspace{-1mm}

\noindent where $edit\_distance$ is the token-level (first parse the code sequence into tokens) Levenshtein distance and $f_{f}^{(i)}$ refers to the given pre-patch function in the hard fork of the $i^{th}$ sample.
Different from AED that measures the absolute edit distances between the generated post-patch function and the ground truth, RED reports the edit distances relative to the one between the pre-patch version and the ground truth.
Thus, if the RED of one approach is less than 1, then the approach is expected to help developers understand where and how to port the patch and reduce the required manual edits.

%% file: tex/experiment_results.tex
\section{Experiment Results}\label{sec:experiment:subsec:results}
\vspace{-0.1cm}

In the experiment, we aim to answer the following two RQs:
\vspace{-0.1cm}
\begin{itemize}[leftmargin=*]
    \item \textbf{RQ1: How effective is \appnamebold compared to baselines for zero-shot patch porting for hard forks?}
    \item \textbf{RQ2: How effective are the key designs of \appnamebold?}
\end{itemize}
\vspace{-0.1cm}

\vspace{-0.2cm}
\subsection{RQ1. The Effectiveness of \appnamebold}~\label{subsec:RQ1}
\vspace{-0.3cm}

\noindent \textbf{Method.} 
To verify the effectiveness of \appname in porting patches, 
we compare its performance with baselines (see Section~\ref{subsec:experiment_setting}) using the test set collected in Section~\ref{subsec:data_preparation_porting}.
Note that the test set only includes patches ported after 2022-07-01, i.e., the date when the corpus used to pretrain~\cite{kocetkov2022stack} and tune (see Section~\ref{subsec:data_preparation_middle}) StarCoder are collected before.
For StarCoder and \appname (i.e., neural models), we investigate with three different settings of length limits (i.e., 2k, 4k, and 8k) to provide a comprehensive evaluation of the performance.
Every LLM has a length limit in its implementation, which restricts the maximum number of tokens (shared by the prompt and the completion) being able to process.
Typical length limits for modern LLMs including 
2k (e.g., GPT-3~\cite{brown2020language}, LLaMA~\cite{touvron2023llama}), 
4k (e.g., GPT-3.5~\cite{openai_models}), 
and 8k for more recent models (e.g., StarCoder~\cite{li2023starcoder}, GPT-4~\cite{openai_models}).
Also, the increase of the input length leads to significant 
(up to quadratic based on the implementation of the transformer attention mechanism)
increase of memory footprint and computational cost.

\input{tex/tables/rq1}

\noindent \textbf{Results.}
Table~\ref{table:RQ1} presents the performance comparisons between \appname and the baselines for zero-shot patch porting for hard forks.
None of the patches in the test set can be directly applied using git-apply or \verb|patch| due to the implementation differences between the source project and the hard fork.
The number of patches that the neural models are able to complete the generation instead of being forced to stop due to the length limit (i.e., the column \emph{Complete} in the table)
increases significantly as we enlarging the length limit from 2k to 8k.
However, even with its maximum limit (i.e., 8k), the StarCoder is still unable to complete the generation of the entire 310 patches in the test set, indicating that some functions could be very lengthy in practice.
This observation highlights the necessities of our proposed reduction module (Section~\ref{subsec:reduction}), which simplifies the function by removing the context less-relevant to the patch.
Compared to StarCoder, \appname is able to complete the generation for more patches, especially under a resource-limit scenario (i.e., 0.8 times more patches with a 2k limit).
For patches that neural models unable to complete the generation, we use the pre-patch version of the function $f_{f}$ as the default output.

\appname@8k (at a length limit of 8k) correctly ports 131 (42.3\%) patches, 9 times more than the one achieved by FixMorph, i.e., 13 (4.2\%) patches.
Regarding edit-based metrics, \appname@8k can significantly reduce the efforts for developers to port the patch by automating 57\% of the manual edits on average (i.e., a RED of 0.43), reducing the average edits to patch a function from 26.60 to 11.98.
FixMorph, on the contrary, introduces 52\% of extra edits (i.e., a RED of 1.52), increasing the average edits to 31.56.
The significant improvement of \appname over FixMorph is mainly due to its ability in understanding the patch semantics, 
thus being able to handle the significant implementation divergences between the source project and the hard fork.
For the motivating example of patch porting presented in Figure~\ref{fig:motivating}, which involves locating and adapting patches with largely changed syntactical structures, 
(see Section~\ref{subsec:motivating_example}).
\appname correctly automates the porting of this patch while FixMorph, as a syntax-based approach, fails to handle such implementation differences.

Compared with the StarCoder,
\appname significantly outperforms the StarCoder in the number of correctly ported patches by 148.3\%, 65.7\%, and 37.9\% under the length limit of 2k, 4k, and 8k, respectively.
Large improvements are also observed on the AED and RED metrics. 
Moreover, \appname even achieves better performances than the StarCoder with a larger length limit, i.e., \appname@2k \emph{vs.} StarCoder@4k and \appname@4k \emph{vs.} StarCoder@8k.
These results show that it is non-trivial to leverage the LLM to effectively automate the patch porting, and verify the validity of our designs in better utilizing LLM's ability for this task (see RQ2 for a detailed discussion). 

\vspace{-3mm}
\find{
{\bf RQ-1:}
\appname significantly outperforms the baselines for zero-shot patch porting in hard forks. 
\appname can effectively help developers understand where and how to port the patches and significantly reduce the required manual edits.
}
\vspace{-0.1cm}

\vspace{-3mm}
\subsection{RQ2. The Key Designs of \appnamebold} \label{subsec:RQ2}
\vspace{-0.1cm}

\noindent \textbf{Method.} 
Our findings in RQ1 have verified that \appname outperforms the baselines by a large margin.
With RQ2, we aim to further verify the effectiveness of the key designs of \appname. 
Specifically, we compare the performances of \appname with two variants, each lacking one of the following key designs:
1)~the design of incorporating the reduction module to slim the function by removing the code snippets less-relevant to the patch (see Section~\ref{subsec:reduction}).
The variant (\appname-r) removes the reduction module (see Figure~\ref{fig:approach_overview}) and directly inputs the original $(f_{s}, f_{s}^{'}, f_{f})$ into the porting module.
2)~the design of a finetuning task to better adapt the LLM to the patch porting task and inject project-specific information (see Section~\ref{subsec:llm_module}).
The variant (\appname-f) removes the finetuning and directly adopts the pretrained \emph{off-the-shelf} LLM.

\input{tex/tables/rq2}

\noindent \textbf{Results.}
Table~\ref{table:RQ2-slice} presents the performance comparisons between \appname and the variant \appname-r.
Since the reduction module slims the input functions by removing the less-relevant contexts, \appname is able to port more patches compared with \appname-r under a given length limit.
To provide an in-depth comparison between \appname and \appname-r, we divide the patches that \appname can complete the generation into following two groups and compare the performance within each group:
\ding{182} patches that both models can complete the generation (column \emph{Share}),
\ding{183} extra patches that only \appname can complete the generation. For these patches, we use the pre-patch version of the function $f_{f}$ as the default output of \appname-r and report the corresponding performance (column \emph{Unique}).

Regarding the \emph{overall} porting performance, \appname increases the performance of \appname-r in the number of correctly ported patches by 111.8\%, 40.5\%, and 13.9\% at the length limits of 2k, 4k, and 8k, respectively.
Large improvements are also observed on AED and RED.
The notable performance improvement is primarily due to \appname's ability to port a larger number of patches under the given length limit (column \emph{Unique} of Table~\ref{table:RQ2-slice}), 
which is enabled by the reduction module as it slims the input functions by removing the less-relevant code snippets.
For example, under the length limit of 2k, \appname is able to complete the generation for 68 more patches than the 82 shared ones with \appname-r, among which \appname correctly ports 38 patches (over 100\% improvement) and reduces the manual edits required to port the patch substantially.

We further compare the performance of \appname and \appname-r regarding the shared patches (column \emph{Share} of Table~\ref{table:RQ2-slice}), which provides us a more comprehensive understanding of the affects introduced by the reduction module.
We find that the incorporation of the reduction module
does not harm the porting performance, 
and could even benefit the performance when the input functions are lengthy, e.g., the RED reduces by 15.7\% and 17.4\% at the length limit of 4k and 8k, respectively.
These observations prove that the code snippets removed by our reduction module are indeed irrelevant for the LLM to port the patch, 
and by removing these irrelevant code snippets, the reduction module can even benefit the patch porting by helping the LLM concentrate on the key contexts related to the patch.
Moreover, the incorporation of the reduction module saves enormous computational cost (including the memory footprint and the run time) for the LLM to process the same patch as the cost is positively correlated to the input length.
The average length after applying the reduction module is 0.70 of the original.

Table~\ref{table:RQ2-slice} presents the performance comparisons between \appname and the variant \appname-f.
Note that we only focus on the porting performances on the patches that models can complete the generation.
The performance improvements brought by the finetuning are substantial and similar across different length limits, e.g., 12.5\%, 12.1\%, and 13.9\% in the number of correctly ported patches under the length limit of 2k, 4k, and 8k, respectively.
These observations verify the effectiveness of introducing the finetuning task to better adapt the LLM to the patch porting task.

\vspace{-0.1cm}
\find{
{\bf RQ-2:}
Both designs improves the performance.
The reduction module slims the input functions, 
enabling \appname to port more patches under the given length limit. 
The introducing of the finetuning task helps the LLM better adapt to the patch porting task.
}
\vspace{-0.1cm}

%% file: tex/tables/rq1.tex
\begin{table}[tbp]
  \centering
  \caption{Performance comparisons between \appname and baselines for zero-shot patch porting for hard forks}
  \vspace{-0.4cm}
  \scalebox{0.8}{
    \begin{tabular}{clcccc}
    \toprule
    \textbf{Length} & \multicolumn{1}{c}{\textbf{Approach}} & \textbf{Complete} & \textbf{Accuracy} & \textbf{AED} & \textbf{RED} \\
    \midrule
    \multirow{4}[2]{*}{/} & Origin & \multirow{4}[2]{*}{/} & 0 (0.0\%) & 26.60  & 1.00  \\
          & git-apply &       & 0 (0.0\%) & /     & / \\
          & patch &       & 0 (0.0\%) & 26.57     & 1.00 \\
          & FixMorph &       & 13 (4.2\%) & 31.56 & 1.52 \\
    \midrule
    \multirow{2}[2]{*}{2k} & StarCoder & 83 (26.8\%) & 29 (9.4\%) & 23.05  & 0.94  \\
          & \appname & 150 (48.4\%) & 72 (23.2\%) & 19.80  & 0.70  \\
    \midrule
    \multirow{2}[2]{*}{4k} & StarCoder & 185  (59.7\%) & 67 (21.6\%) & 18.76  & 0.83  \\
          & \appname & 242 (78.1\%) & 111 (35.8\%) & 14.45  & 0.51  \\
    \midrule
    \multirow{2}[2]{*}{8k} & StarCoder & 255 (82.3\%) & 95 (30.6\%) & 14.30  & 0.67  \\
          & \appname & 290 (93.5\%) & 131 (42.3\%) & 11.98  & 0.43  \\
    \bottomrule
    \end{tabular}}
    \vspace{-0.5cm}
  \label{table:RQ1}%
\end{table}

%% file: tex/tables/rq2.tex
\begin{table*}[htbp]
  \centering
  \caption{Comparisons of \appnamebold with variant \appnamebold-r}
  \vspace{-0.4cm}
  \scalebox{0.8}{
    \begin{tabular}{clcccc|cccc|cccc}
    \toprule
    \multirow{2}[2]{*}{\textbf{Length}} & \multirow{2}[2]{*}{\textbf{Approach}} & \multicolumn{4}{c|}{\textbf{Overall}} & \multicolumn{4}{c|}{\textbf{Share}} & \multicolumn{4}{c}{\textbf{Unique}} \\
          &       & \textbf{Complete} & \textbf{Accuracy} & \textbf{AED} & \textbf{RED} & \textbf{Complete} & \textbf{Accuracy} & \textbf{AED} & \textbf{RED} & \textbf{Complete} & \textbf{Accuracy} & \textbf{AED} & \textbf{RED} \\
    \midrule
    \multirow{2}[2]{*}{2k} & \appname-r & 82 (26.5\%) & 34 (11.0\%) & 22.76  & 0.85  & \multirow{2}[2]{*}{82} & 34    & 7.95  & 0.44  & 0     & 0     & 19.50  & 1.00  \\
          & \appname & 150 (48.4\%) & 72 (23.2\%) & 19.80  & 0.70  &       & 34    & 8.22  & 0.43  & 68    & 38    & 5.68  & 0.30  \\
    \midrule
    \multirow{2}[2]{*}{4k} & \appname-r & 185 (59.7\%) & 79 (25.5\%) & 18.37  & 0.71  & \multirow{2}[2]{*}{185} & 79    & 11.52  & 0.51  & 0     & 0     & 23.58  & 1.00  \\
          & \appname & 242 (78.1\%) & 111 (35.8\%) & 14.45  & 0.51  &       & 80    & 10.51  & 0.43  & 57    & 31    & 5.54  & 0.20  \\
    \midrule
    \multirow{2}[2]{*}{8k} & \appname-r & 255 (82.3\%) & 115 (37.1\%) & 14.42  & 0.56  & \multirow{2}[2]{*}{255} & 115   & 11.95  & 0.46  & 0     & 0     & 21.63  & 1.00  \\
          & \appname & 290 (93.5\%) & 131 (42.3\%) & 11.98  & 0.43  &       & 117   & 10.88  & 0.38  & 35    & 14    & 7.83  & 0.45  \\
    \bottomrule
    \end{tabular}}
    \vspace{-0.3cm}
  \label{table:RQ2-slice}%
\end{table*}%

\begin{table}[htbp]
  \centering
  \caption{Comparisons of \appnamebold with variant \appnamebold-f}
  \vspace{-0.4cm}
  \scalebox{0.8}{
    \begin{tabular}{clcccc}
    \toprule
    \textbf{Length} & \multicolumn{1}{c}{\textbf{Approach}} & \textbf{Complete} & \textbf{Accuracy} & \textbf{AED} & \textbf{RED} \\
    \midrule
    \multirow{2}[2]{*}{2k} & \appname-f & \multirow{2}[2]{*}{150} & 64 (42.7\%) & 7.83  & 0.50  \\
          & \appname &       & 72 (48.0\%) & 7.07  & 0.37  \\
    \midrule
    \multirow{2}[2]{*}{4k} & \appname-f & \multirow{2}[2]{*}{242} & 99 (40.9\%) & 9.48  & 0.47  \\
          & \appname &       & 111 (45.9\%) & 9.34  & 0.37  \\
    \midrule
    \multirow{2}[2]{*}{8k} & \appname-f & \multirow{2}[2]{*}{290} & 115 (39.7\%) & 10.35  & 0.46  \\
          & \appname &       & 131 (45.2\%) & 10.51  & 0.39  \\
    \bottomrule
    \end{tabular}}
    \vspace{-0.4cm}
  \label{tab:addlabel}%
\end{table}%

%% file: tex/discussion.tex
\vspace{-1 mm}
\section{Discussion} \label{sec:discussion}
\vspace{-1 mm}
In this section, we discuss 
the effectiveness of \appname in porting security patches,
the generalizability of \appname regarding different hard fork pairs,
and the threats to validity.

\vspace{-0.1cm}
\subsection{Porting Security Patches} \label{subsec:discuss_security}
\vspace{-0.1cm}
We evaluate the effectiveness of \appname in porting security patches using CVE fixes ported from Vim to Neovim after 2022-07-01.
There are 41 CVE fixes that only involves modifications of a single function.
\appname can correctly port nearly half (i.e., 20) of the patches.
Besides, \appname can automate 78\% of the edits on average (i.e., a RED of 0.22) required for developers to manually port the patch. 

Besides, we further conduct a case study with 10 CVE fixes to investigate the performance of \appname in porting patches that modify multiple functions.
For each CVE fix, we leverage \appname to port the patch on a function-wise basis.
Table~\ref{table: security_patch_multi} shows the statistics of the selected CVE fixes 
and the porting results.
\appname correctly ports three CVE fixes, and is expected to correctly port the patch to over half of the modified functions (1.7/2.9) involved in a CVE fix on average.
\appname can help developers in porting patches with multi-function changes by significantly reducing the required manual edits, i.e., reducing the average edits to port a CVE fix from 60.2 to 18.8.
The above results verify the effectiveness of \appname in porting the patches with multi-function changes.

\vspace{-0.2cm}
\begin{table}[tbp]
  \centering
  \caption{Case study of porting multi-function change patches}
  \vspace{-0.35cm}
    \scalebox{0.8}{
    \begin{threeparttable}
    \begin{tabular}{cccccc}
    \toprule
    CVE ID & \tabincell{c}{Vim\\Patch} & \tabincell{c}{Neovim\\Patch} & Accuracy\tnote{1} & \tabincell{c}{AED\\(Origin)} & RED \\
    \midrule
    CVE-2022-1725 & b62dc5e & ddd69a6 & 3/3   & 0 (62) & 0 \\
    CVE-2022-1897 & 338f1fc & df4c634 & 4/5   & 3 (110) & 0.03 \\
    CVE-2022-1942 & 71223e2 & 0612101 & 2/4   & 13 (69) & 0.19 \\
    CVE-2022-2522 & 5fa9f23 & 8e67af1 & 2/4   & 4 (50) & 0.08 \\
    CVE-2022-2598 & 4e677b9 & 88ed332 & 1/2   & 4 (36) & 0.11 \\
    CVE-2022-3016 & 6d24a51 & f72ae45 & 3/3   & 0 (25) & 0 \\
    CVE-2022-3296 & 96b9bf8 & 0cb9011 & 0/2   & 121 (147) & 0.82 \\
    CVE-2022-3324 & 8279af5 & cd9ca70 & 2/2   & 0 (18) & 0 \\
    CVE-2022-3352 & ef97632 & 6bc2d6b & 0/2   & 10 (40) & 0.25 \\
    CVE-2023-2426 & caf642c & ab7dcef & 0/2   & 33 (45) & 0.73 \\
    \midrule
    Avg.  & -     & -     & 1.7/2.9 & 18.8 (60.2) & 0.22 \\
    \bottomrule
    \end{tabular}
    \begin{tablenotes}
      \footnotesize
      \item[1] The Accuracy for each CVE fix is presented in the form of \{\#correctly patched functions\}/\{\#total functions changed by the patch\}.
    \end{tablenotes}
    \end{threeparttable}}
    \vspace{-0.3cm}
  \label{table: security_patch_multi}%
\end{table}%

\input{tex/discussion_anotherpair}

\vspace{-0.1cm}
\subsection{Threats to Validity} \label{subsec:threats}
\vspace{-0.1cm}

\noindent \textbf{Threats to internal validity} refer to the experiment biases and errors.
Threat related to our approach is that in our reduction module, when mapping the removable code segments between the source project and the hard fork, we generally rely on the text similarity between two code segments.
It can fail if the hard fork changes the namespaces and implementations significantly. 
More precise mapping techniques such as semantic clone detection are required in such cases.
Another threat regarding evaluation is that we regard the developer ported patch as the ground truth to calculate the metrics. 
However, a patch that is syntactically different but semantically equivalent to the developer ported patch should also be considered as a correct porting.
Thus, our evaluation may under estimate the porting performance.
Nevertheless, it is challenging and can be error-prone to determine whether a patch is semantically equivalent to the developer ported patch.

\noindent \textbf{Threats to external validity} refer to the generalizability of our approach.
We evaluate \appname using 310 patches with single-function change.
We conduct a case study to show that \appname can also port patches that involves modifications of multiple functions.
However, more evaluations are needed to investigate the effectiveness of \appname in porting patches where the functions from the source project and the hard fork do not conform to a one-to-one mapping.
For example, changes of multiple functions from the source project should be ported into one function in the hard fork.
Besides, though we evaluate \appname on two additional hard fork pairs (i.e., Cpython-Cinder, FreeBSD-OpenBSD) apart from the Vim-Neovim (see Section~\ref{subsec:discuss_generalizability}), the evaluation sets for these two additional pairs are relatively small. 
This is because these two hard fork pairs do not standardize patch porting practice (e.g., lack unified documentation),
making it difficult and time-consuming for us to collect the ported patch pairs.
Future works are required to collect larger datasets and include more hard fork pairs to more comprehensively evaluating the generalizability of \appname.
Furthermore, we only investigate the effectiveness of \appname with StarCoder as the underlying LLM. 
Although StarCoder is one of the most advanced open-source code LLMs when we conduct our experiments, code LLMs are undergoing rapid updates and more advanced LLMs (e.g., CodeLlama~\cite{roziere2023code}) are constantly being introduced. 
Future works should integrate \appname with more advanced code LLMs to investigate its generalizability across different underlying LLMs.

%% file: tex/discussion_anotherpair.tex
\subsection{Porting Patches of Other Hard Fork Pairs} \label{subsec:discuss_generalizability}
\vspace{-0.1cm}

\begin{table}[tbp]
  \centering
  \caption{Patch porting performance of other hard fork pairs}
  \vspace{-0.35cm}
  \scalebox{0.8}{
    \begin{tabular}{llccc}
    \toprule
    Hard Fork Pair & \multicolumn{1}{c}{Approach} & Accuracy & AED (Origin) & RED \\
    \midrule
    \multirow{2}[2]{*}{Cpython-Cinder} & FixMorph & 1/13 (7.7\%) & 27.85 (35.77) & 0.77 \\
          & \appname & 8/13 (61.5\%) & 9.46 (35.77) & 0.19 \\
    \midrule
    \multirow{2}[2]{*}{FreeBSD-OpenBSD} & FixMorph & 1/7 (14.3\%) & 16.29 (18.14) & 0.85 \\
          & \appname & 5/7 (71.4\%) & 8.0 (18.14) & 0.17 \\
    \bottomrule
    \vspace{-0.8cm}
    \end{tabular}}
  \label{table:discussion_generalizability}%
\end{table}%

To verify the generalizability of \appname, 
we evaluate its porting performance with 
another two well-known hard fork pairs, i.e., Cpython-Cinder and FreeBSD-OpenBSD.
Both two hard forks do not standardize the patch porting practice (e.g., lack unified documentation). 
We first filter potential ported patches by searching for commit hash or pull request id of the source project mentioned in the commit message, and then manually validate them and extract the patched function pairs.
We only consider patches ported after 2022-07-01 for evaluation. 
We collect 13 and 7 patches for Cpython-Cinder and FreeBSD-OpenBSD, respectively.
We do not finetune \appname again using commits from these two hard fork pairs.
Table~\ref{table:discussion_generalizability} presents the porting performance of FixMorph and \appname on the two hard fork pairs.
\appname outperforms FixMorph by a large margin.
For both hard fork pairs, \appname can correctly port over 60\% of the investigated patches, and reduce over 80\% (i.e., with RED $< 0.2$) of the manual edits.
These results verify the generalizability of \appname regarding different hard fork pairs.

%% file: tex/relatedwork.tex
\vspace{-0.1cm}
\section{Related Work} \label{sec:relatedwork}
In this section, we describe two aspects of the related work:

\noindent \textbf{Hard Forks.}
We follow Zhou~\emph{et al.}~\cite{zhou2020has, zhou2019fork, zhou2020thesis} to distinguish between two forms of forks by referring the traditional one with the intention of code reuse as \emph{hard forks}, and the new one with the intention of collaborative development as \emph{social forks}.
Our focus in this paper is to automate the patch propagation across a family of hard forks (i.e., independent projects with a relation of code reuse).
There are a bunch of studies focus on investigating the motivations and outcomes of hard forks~\cite{robles2012comprehensive, jiang2017and}, as well as its pros and cons to the software development~\cite{ray2012case, businge2022reuse}.
Recently, Zhou~\emph{et al.}~\cite{zhou2020has} conduct an empirical study to revisit the general practice of hard forks on GitHub.
Businge~\emph{et al.}~\cite{businge2022reuse} empirically investigate the maintenance practices among the hard fork families in three software ecosystems (e.g., Android, .NET, and JavaScript).
Both Zhou~\emph{et al.} and Businge~\emph{et al.} conclude that 
current tools (e.g., \verb|git apply|) fail to handle the divergence between hard forks, and the lack of integration support tools has resulted in limited coordination and duplicate efforts.
Ren~\emph{et al.}~\cite{ren2019automated} conduct an empirical study and find that a non-negligible portion (20.5\%) of patches need to be ported across forked projects.
Different from the existing works that are mainly focus on understanding the general behaviour of hard forks (e.g., motivations), we highlight the flaws in the current patch porting practice across hard forks (especially from the security perspective) and further propose an approach to automate the porting process.

\noindent \textbf{Patch Transplantation.}
There are several works that try to transplant patch from one codebase to another.
Automated program transformation~\cite{bader2019getafix, SLA_phoenix} approaches infer transformation rules from the exiting patches, and then apply the inferred rules to new codebases.
However, these approaches need to learn the transformation rules from multiple similar patches.
FixMorph~\cite{shariffdeen2021automated} backports the patch from the mainline version of Linux to older versions.
It abstracts the patch into transformation rules at the syntax level. 
It leverages the syntactic structure to find the patch locations and apply the abstracted transformation rule.
PatchWeave~\cite{shariffdeen2020automated} transplant a patch from a donor program to fix the vulnerability. 
It relies on test cases and concolic execution to locate the patch insertion point and performs namespace translations.
There are two drawbacks that limits the application of these methods to port patches for hard forks:
1) they can not deal with complicated adaptions, e.g., the syntactic structure of the patch needs to be modified,
2) they rely on extra information (e.g., multiple similar patches, available test cases) apart from the reference patch from the source project, which are not always available in reality.
In this work, we leverage LLM's superior understanding of code semantics and strong performance under zero-shot setting to automate patch porting for hard forks.

%% file: tex/conclusion.tex
\vspace{-1mm}
\section{Conclusion and Future Work} \label{sec:conclusion}
In this paper, we take the first step to automate the patch porting for hard forks.
We propose a LLM-based approach (namely \appname) to port the patch on a function-wise basis.
\appname is composed of a reduction module and a porting module.
The reduction module first slims the input functions by removing code snippets less relevant to the patch, enabling \appname to port more patches under the given length limit.
The porting module effectively leverages a LLM to apply the patch from the reference project to the target project by 
1) enabling efficient in-context learning with a specially designed prompt template,
2) effectively adapting the LLM to the patch porting and injecting project-specific knowledge with the proposed instruction-tuning based training task.
The evaluation results on 310 Neovim patches ported from Vim demonstrate that \appname outperforms the baselines substantially.
\appname correctly ports 131 (42.3\%) patches, and can help developers understand where and how to port the patch by reducing 57\% of the manual edits.

In the future, we plan to integrate other LLMs with our approach and apply it to other projects and programming languages.